# Single Pixel MEMS Spectrometer using Electrothermal Tunable Grating

JAEHUN JEON, JUNG-WOO PARK, GI BEOM KIM, MYEONG-SU AHN, AND KI-HUN JEONG[*]

Department of Bio and Brain Engineering, Korea Advanced Institute of Science and Technology (KAIST), 291 Daehak-ro, Yuseong-gu, Daejeon 34141, Republic of Korea
KAIST Institute for Health Science and Technology (KIHST), KAIST, 291 Daehak-ro, Yuseong-gu, Daejeon, 34141, Republic of Korea

*kjeong@kaist.ac.kr

**Abstract:** Miniaturized spectrometers are widely used for non-destructive and on-field spectral analysis. Here we report a tunable grating-based MEMS spectrometer for visible to near-infrared (VIS-NIR) spectroscopy. The MEMS spectrometer consists of a spherical mirror and an electrothermally actuated tunable grating. The spectrometer detects the dispersed spectral signal with a single-pixel detector by tilting the diffraction grating. The large tilting angle from electrothermal actuation and highly dispersive diffraction grating improves the spectral range and resolution, respectively. The MEMS spectrometer was fully packaged within 1.7 $cm^3$ and provides a measurable spectral range up to 800 nm with an average 1.96 nm spectral resolution. This miniaturized single-pixel spectrometer can provide diverse applications for advanced mobile spectral analysis in agricultural, industrial, or medical fields.

## 1. Introduction

Spectroscopy facilitates rapid and non-destructive sample analysis by measuring spectra from absorbance [1], reflectance [2], or Raman scattering [3, 4]. Therefore, miniaturized spectrometers are widely used for on-field sample analysis in the medical field such as hemoglobin measurement [5] as well as in industrial and agricultural applications such as plastic classification [6, 7] and food quality checking [8]. Further miniaturization of the spectrometer provides high utility in mobile applications, however, the trade-off between the optical path length, which restricts miniaturization, and the spectral resolution [9] impedes the accurate on-field spectral analysis using the miniaturized dispersive spectrometer.

The dispersive spectrometer generally comprises dispersive elements, collimating and focusing optics, and detector. Three design parameters mainly determine the spectral performance of the system: optical path length, dispersion angle, and slit width. For instance, long optical path length and high dispersion angle improve spectral resolution. In addition, narrow slit width also improves the spectral resolution, however, suffers from the low intensity of radiation. The folded spectrometer configuration using metasurface is an example, which improves spectral resolution by extending optical path length in limited physical volume [10]. However, numerous miniaturized dispersive spectrometers achieve high spectral performance by applying highly dispersive elements [11, 12] and narrow slits [13] due to the high fabrication complexity of the metasurface. Those systems are affected by the optical aberration from the curved focusing plane known as the Rowland circle and suffer from wavelength-dependent performance [11]. In addition, the sensitivity and resolution are in a trade-off relation due to the narrow slits. These limitations hinder accurate low-level light analysis with the miniaturized spectrometer.

Single-pixel spectroscopy provides a breakthrough in low-level light analysis by using a highly sensitive detector such as photomultiplier tube, avalanche photodiode [14], and superconducting nanowire single-photon detector [15]. In addition, the system is less affected by optical aberration such as Rowland circles and has benefits in miniaturization and power consumption since the cooling system is relatively unnecessary, unlike CCD line sensors [14]. Several methods for single-pixel spectroscopy were presented by applying MEMS-based Fourier transform spectroscopy [16, 17], dispersive Fourier transform spectroscopy [18], compressive sensing using digital micromirror device (DMD) [19], and MEMS tunable grating [20, 21]. Especially, MEMS tunable gratings provide the integrated configurations of dispersive element and MEMS actuator, which is favorable for the miniaturization of the entire system. However, previously reported electrostatic [20] and electromagnetic [21] devices for near-infrared spectral range still require large strokes with highly dispersive elements for improved spectral resolution and a wide spectral range.

Here we report a single-pixel MEMS spectrometer using electrothermal tunable grating. Figure 1a shows the schematic illustration of the MEMS spectrometer. The MEMS spectrometer consists of a tunable grating and a spherical mirror for collimating and focusing elements. The optical elements including reflective diffraction grating, electrothermal actuator, and slits are integrated into a single chip. The tunable grating facilitates the detection of the dispersed beam with a single-pixel photodiode by tilting as shown in Figure 1b. The spectral signals in the time unit are converted into a wavelength unit by applying the applied voltage to the tilting angle and tilting angle to the wavelength calibration process. In addition, a highly dispersive diffraction grating of the device improves spectral resolution and electrothermal actuation implements a large tilting angle for a wide spectral range. The MEMS spectrometer only detects the beams with the same optical path, therefore, provides uniform spectral resolution in the whole spectral range. The fully packaged MEMS spectrometer provides a visible to near-infrared (VIS-NIR) spectral range up to 800 nm with an average 1.96 nm spectral resolution within 1.7 cm$^3$ volume.

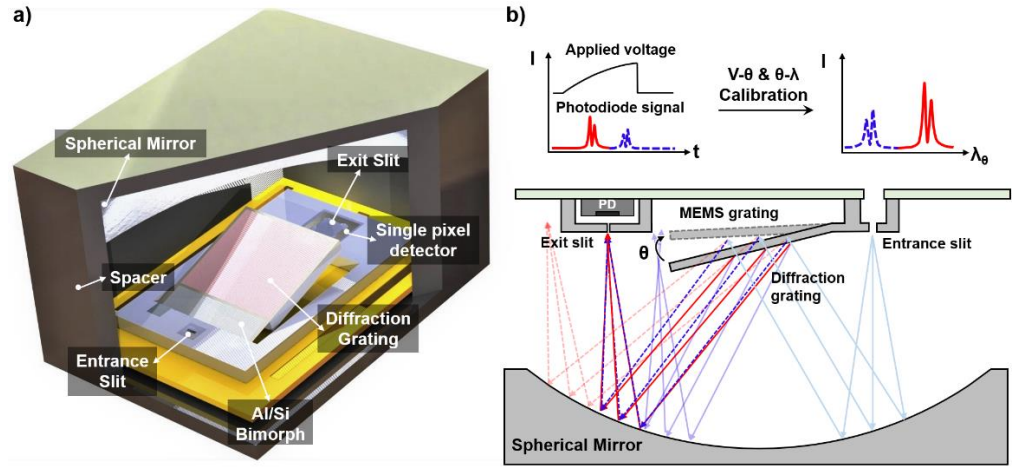

Figure 1 a) Schematic illustrations of the MEMS spectrometer using electrothermal tunable grating, b) and cross-section of the spectrometer with calibration method. The MEMS spectrometer consists of a spherical mirror and a tunable grating. The MEMS spectrometer facilitates the detection of the dispersed spectral signal with a single-pixel detector by tilting. The collected time-unit spectral signals are converted into a wavelength unit by applying the applied voltage to the tilting angle and the tilting angle to the wavelength calibration process.

## 2. Microfabrication of electrothermal tunable grating

The electrothermal tunable grating was fabricated on a 6-inch SOI wafer (silicon-on-insulator wafer, top Si: 2.5 *μm*, buried oxide layer: 1 *μm*, bottom Si: 430 *μm*). Figure 2a shows the microfabrication procedure of the tunable grating. First, 200 nm low-stress Silicon nitride was deposited via low-pressure chemical vapor deposition (LPCVD) for insulation and diffraction grating pattern. The deposited Silicon nitride was etched by reactive ion etching to pattern the insulation and diffraction grating pattern. The top Silicon layer works as etch stop layer. The diffraction grating has a 1.25 *μm* pitch with a 0.4 duty cycle for improved spectral resolution in the VIS-NIR range and fabrication precision in stepper photolithography. A 100 nm aluminum film was deposited and patterned for reflective diffraction grating and slits by thermal evaporation and wet etching, respectively. A 1 *μm* thick aluminum film was deposited and patterned by lift-off for Si/Al bimorph having a repeated line pattern with a height of 1,000 *μm*, a width of 14 *μm* at a period of 20 *μm*. Top Silicon was etched by deep reactive ion etching (DRIE) for slit patterns. The exit slit has a 10 *μm* width, and the entrance beam passes through a core with a diameter of 3 *μm*. Finally, the backside silicon opening and silicon rim defining were conducted by using DRIE, and the buried oxide layer was etched in buffered oxide etchant (BOE). The fabricated devices are tilted by the residual stress from the aluminum line pattern. In addition, the hollow silicon rim structure protects the diffraction grating from the deformation, which is caused by residual stress of Al thin film on the substrate [22]. Figure 2b shows the optical image of the microfabricated tunable grating and 6-inch wafer-level fabrication result (right bottom). Figure 2c and Figure 2d show the SEM image of the microfabricated diffraction grating and microscopy image of the silicon rim structure, respectively.

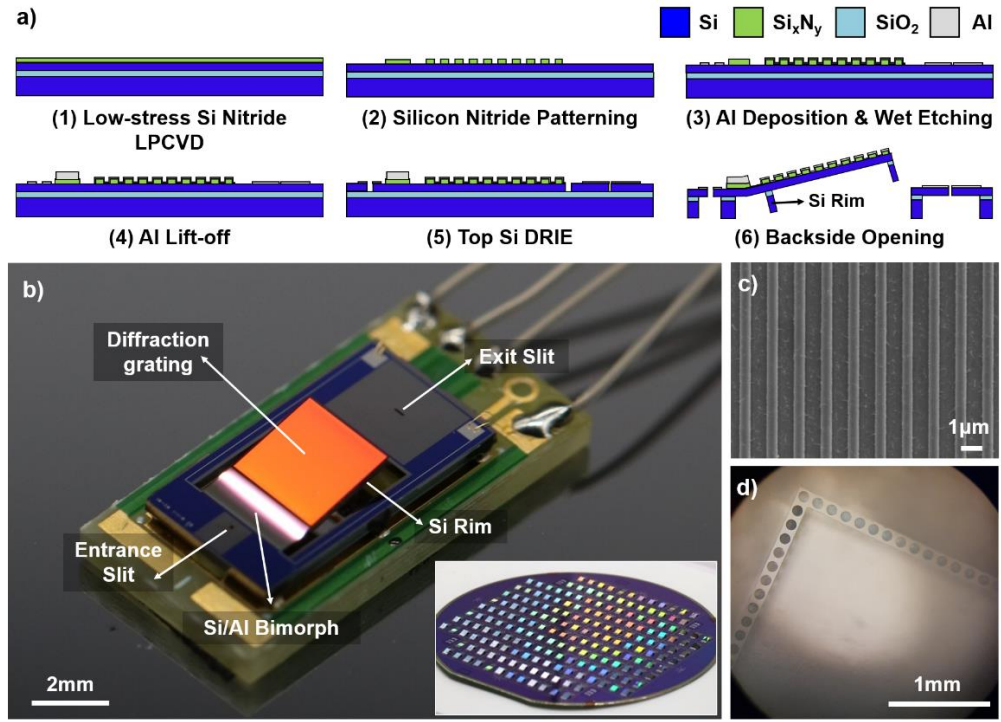

Figure 2 a) Microfabrication procedure of electrothermal tunable grating. The tunable grating is fabricated on a 6-inch SOI wafer. b) Optical image of microfabricated tunable grating, and 6-inch wafer-level fabrication result (right bottom). The tunable grating consists of a reflective diffraction grating, Si/Al bimorph, and slits. c) SEM image of microfabricated diffraction grating on the device and d) microscopy image of Si rim. The hollow Si rim is simply fabricated by DRIE and protects the diffraction grating from the deformation, which is caused by residual stress of Al thin film on the substrate.

## 3. Characterization of tunable grating and calibration method

The tunable grating is electrothermally actuated and Figure 3a shows the measured and calculated tilting angle along to the applied DC voltage in static mode. The tilting angle was calculated by applying finite element methods (FEM, COMSOL Multi-physics® ver. 5.5). In addition, the measured resonant frequency of the device is 92 $Hz$. The tunable grating tilts 14° at 7 $V_{DC}$ applied voltage and each tilting angle corresponds to the specific wavelength. For instance, high and low angles correspond to the NIR and VIS range, respectively. Figure 3b shows applied voltage and measured laser spectra on a time scale. The applied voltage consists of a function of square root with 5s duration and 0 $V_{DC}$ for the cooling process. The spectrum in the time scale is converted into the wavelength scale with the calibration between time-applied voltage and applied voltage-wavelength. Each square represents the corresponding applied voltage to the specific wavelength and $0^{th}$ order beam. Figure 3c shows relations between applied DC voltage and wavelength (red dotted line). The calibration process is performed in the relative coordinate system and the reference is the $0^{th}$ order diffraction beam. In addition, differences between squares of the applied voltage corresponding to the $1^{st}$ and $0^{th}$ order diffraction beam are used for the calibration. The values are linear with the wavelength as shown in Figure 3c (blue dotted line). The data points represent the average of 30 different experimental results and the standard deviation of the calibration results is less than 0.6 nm. The MEMS spectrometer provides VIS-NIR spectral range up to 800 nm at 7 $V_{DC}$ applied voltage.

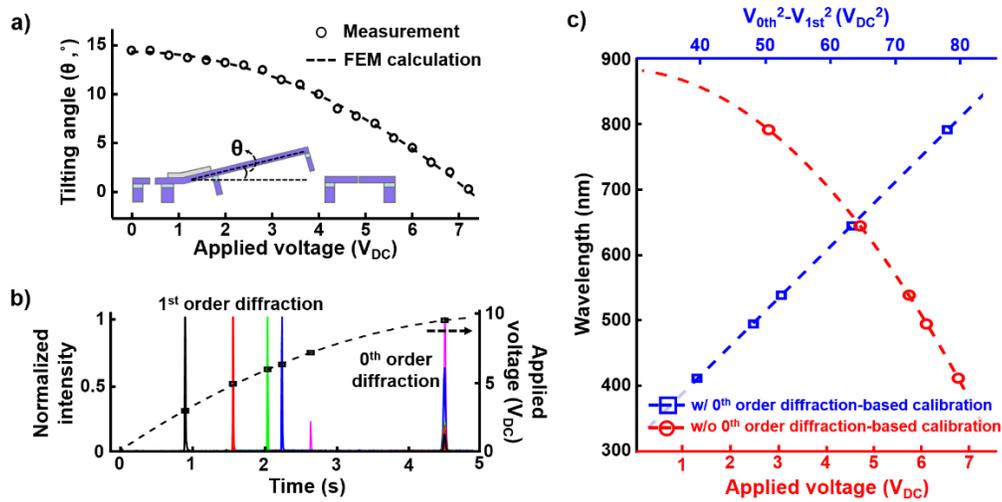

Figure 3 a) Measured and calculated tilting angle of the tunable grating along to the applied voltage in static mode. b) Applied DC voltage and measured laser spectra on a time scale. Each color represents a different wavelength (magenta: 405nm, blue: 488 nm, green: 532 nm, red: 638 nm, black: 785 nm). The squares represent the corresponding applied DC voltage to the wavelength of the $1^{st}$ order diffraction beam and the $0^{th}$ order diffraction beam. c) Calibration graph between applied voltage and wavelength. Differences between squares of applied voltage corresponding to the $1^{st}$ and the $0^{th}$ order diffraction beam are linear with wavelength.

## 4. Fully packaged MEMS spectrometer

The MEMS spectrometer was fully packaged within 1.7cm$^3$. The tunable grating and photodiode are bonded to the printed circuit board (PCB) and stacked vertically with a spherical mirror (Edmund optics, 43-462) using a 3d printed spacer. The distance between the spherical mirror and the tunable grating is equal to half of the radius of curvature of the spherical mirror, which is 6mm. Figure 4a and Figure 4b show PCB-bonded tunable grating with a photodiode and fully packaged spectrometer, respectively. Figure 4c shows the measured spectra of 5 different lasers with the fully packaged MEMS spectrometer. In addition, Figure 4d shows the measured spectral resolution and diffraction efficiency along the wavelength. The average spectral resolution in the spectral range of 400 nm to 800 nm is measured as 1.96 nm. The diffraction efficiencies of the fabricated rectangular grating range from 15 % to 50 %. The spectral resolution and diffraction efficiency can be improved by using short pitch blazed grating. Finally, Rhodamine 6G (R6G) fluorescence spectrum is measured by using the MEMS spectrometer. Figure 4e shows the schematic illustration of the optical setup for fluorescence measurement. A 532 nm laser with 1.8 mW power was used as an excitation beam and the long pass dichroic mirror (Semrock, FF555-Di03) has a cut-on wavelength at 555 nm. Figure 4f shows the measured fluorescence emission spectra of 300 μM R6G solution using the MEMS spectrometer and a commercial spectrometer (Hamamatsu Inc., C10082CAH). The MEMS spectrometer clearly captures the emission tendency of R6G. The signal-to-noise ratio (SNR) of the MEMS grating can be improved by designing a low-noise current to voltage circuit.

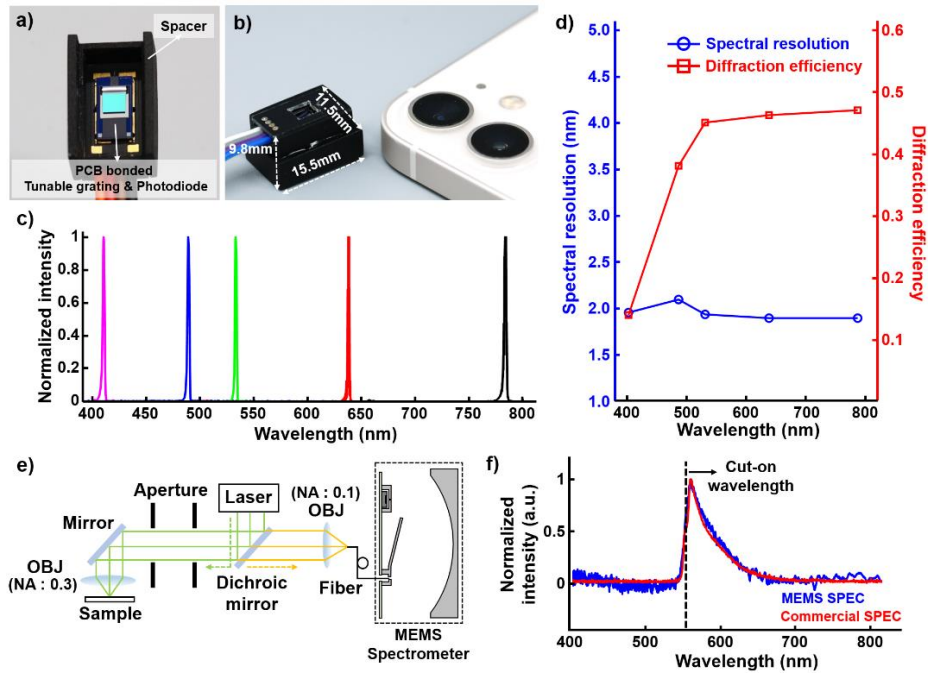

Figure 4 Optical images of a) PCB-bonded tunable grating and photodiode with spacer, and b) fully packaged MEMS spectrometer. c) Measured spectra of the lasers for performance evaluation. d) Measured spectral resolution and diffraction efficiency of the fully packaged spectrometer. e) Schematic illustration of the optical setup for fluorescence spectrum measurement. f) Measured fluorescence emission spectrum of 300 μM R6G solution using the MEMS spectrometer (blue line) and a commercial spectrometer (red line).

## 5. Conclusion

In summary, we have demonstrated the single-pixel VIS-NIR MEMS spectrometer using electrothermal tunable grating for improved spectral resolution and wide range. The tunable grating consists of the reflective diffraction grating, Si/Al bimorph, and slits. The tunable grating provides a 14 ° tilting angle with 7 $V_{DC}$ applied voltage in static mode. The MEMS spectrometer was fully packaged within 1.7cm$^3$ and provides a measurable spectral range up to 800 nm and an average 1.96 nm spectral resolution. The MEMS spectrometer can be used for low-level light measurements such as Raman or surface-enhanced Raman spectroscopy by applying highly sensitive single-pixel detectors. This miniaturized spectrometer will provide diverse applications for advanced mobile spectral analysis in agricultural, industrial, or medical fields.

**Disclosures** The authors declare no conflicts of interest.

**Data availability** Data underlying the results presented in this paper are not publicly available at this time but may be obtained from the authors upon request.

**Acknowledgments** This work was supported by the Korea Medical Device Development Fund (KDMF_PR_20200901_0074) and the National Research Foundation of Korea (NRF) funded by the Ministry of Science ICT & Future Planning (2021R1A2B5B03002428).